\documentclass[aps,prb,twocolumn,superscriptaddress,amssymb,reprint,showpacs]{revtex4-1}

\usepackage[ansinew]{inputenc}
\usepackage[T1]{fontenc}
\usepackage{ae,aecompl}
\usepackage[english]{babel}
\usepackage{graphicx}
\usepackage{hyperref}
\usepackage{color}
\usepackage{multirow}
\usepackage{natbib}
\usepackage{textcomp}

\begin{document}

\author{F. Arnold}
\affiliation{Max Planck Institute for Chemical Physics of Solids, 01187 Dresden, Germany}
\author{M. Naumann}
\affiliation{Max Planck Institute for Chemical Physics of Solids, 01187 Dresden, Germany}
\affiliation{Physik-Department, Technische Universit\"at M\"unchen, 85748 Garching, Germany}
\author{H. Rosner}
\affiliation{Max Planck Institute for Chemical Physics of Solids, 01187 Dresden, Germany}
\author{N. Kikugawa}
\affiliation{National Institute for Materials Science, Tsukuba, Ibaraki 305-0003, Japan}
\author{D. Graf}
\affiliation{National High Magnetic Field Laboratory, Florida State University, Tallahassee, FL 32310, USA}
\author{L. Balicas}
\affiliation{National High Magnetic Field Laboratory, Florida State University, Tallahassee, FL 32310, USA}
\author{T. Terashima}
\affiliation{National Institute for Materials Science, Tsukuba, Ibaraki 305-0003, Japan}
\author{S. Uji}
\affiliation{National Institute for Materials Science, Tsukuba, Ibaraki 305-0003, Japan}
\affiliation{Graduate School of Pure and Applied Sciences, University of Tsukuba, Tsukuba 305-8577, Japan}
\author{H. Takatsu}
\affiliation{Department of Physics, Tokyo Metropolitan University, Tokyo 192-0397, Japan}
\affiliation{present adress: Department of Energy \& Hydrocarbon Chemistry, Graduate School of Engineering, Kyoto University, Kyoto 615-8510 Japan}
\author{S. Khim}
\affiliation{Max Planck Institute for Chemical Physics of Solids, 01187 Dresden, Germany}
\author{A.P. Mackenzie}
\affiliation{Max Planck Institute for Chemical Physics of Solids, 01187 Dresden, Germany}
\affiliation{Scottish Universities Physics Alliance, School of Physics and Astronomy, University of St. Andrews, St. Andrews, Fife KY16 9SS, UK}
\author{E. Hassinger}
\thanks{correspondence should be addressed to elena.hassinger@cpfs.mpg.de} 
\affiliation{Max Planck Institute for Chemical Physics of Solids, 01187 Dresden, Germany}
\affiliation{Physik-Department, Technische Universit\"at M\"unchen, 85748 Garching, Germany}

\title{The Fermi surface of PtCoO$_2$ from quantum oscillations and electronic structure calculations}
\date{\today}

\begin{abstract}
The delafossite series of layered oxides include some of the highest conductivity metals ever discovered.  Of these, PtCoO$_2$, with a room temperature resistivity of 1.8 \textmu$\Omega$cm for in-plane transport, is the most conducting of all.  The high conduction takes place in triangular lattice Pt layers, separated by layers of Co-O octahedra, and the electronic structure is determined by the interplay of the two types of layer.  We present a detailed study of quantum oscillations in PtCoO$_2$, at temperatures down to 35 mK and magnetic fields up to 30 T.  As for PdCoO$_2$ and PdRhO$_2$, the Fermi surface consists of a single cylinder with mainly Pt character, and an effective mass close to the free electron value.  Due to Fermi-surface warping, two close-lying high frequencies are observed. Additionally, a pronounced difference frequency appears.  By analysing the detailed angular dependence of the quantum-oscillation frequencies, we establish the warping parameters of the Fermi surface.  We compare these results to the predictions of first-principles electronic structure calculations including spin-orbit coupling on Pt and Co and on-site correlation $U$ on Co, and hence demonstrate that electronic correlations in the Co-O layers play an important role in determining characteristic features of the electronic structure of PtCoO$_2$.

\end{abstract}

\maketitle
\section{Introduction}
The delafossite series of oxides, with the general formula ABO$_2$, are based on stacks of triangular-coordinated layers with a three-unit repeat resulting in point-group symmetry $R\bar{3}m$ for the most common $3R$ structural isomorph.\cite{Prewitt71}  They host a large number of AB combinations, with A site atoms including Pt, Pd, Ag and Cu, and B site transition metals including Cr, Ge, Co, Ni, and Rh in an octahedral co-ordination with oxygen.\cite{Shannon71}  The formal valence of the transition metal is 3+, sometimes (e.g. in the case of Cr) resulting in a local magnetic moment but in other cases, notably Co$^{3+}$ in its low-spin 3$d^6$ configuration, in non-magnetic ions. Most delafossites are insulators (often magnetic) and semimetals, but five (PdCoO$_2$, PdCrO$_2$, PdRhO$_2$, PtCoO$_2$ and AgNiO$_2$) are known to be metals. \cite{Mackenzie17}  In AgNiO$_2$, the conduction is due to Ni-O states, and occurs in a complex phase diagram in which conducting phases are interleaved with charge-ordered insulators.\cite{Coldea14, Coldea09a}  In contrast, in the Pd- and Pt- based materials the conduction results dominantly from 4$d$-5$s$ or 5$d$-6$s$ states of Pd or Pt respectively, leading to large bandwidths and almost free electron conduction in these quasi-two-dimensional metals.  The conductivity is extremely high, with the mean free path deduced from electrical transport measurements rising from nearly 1000\,\AA\,\,at room temperature to ten microns or more below 10 K. \cite{Takatsu07, Takatsu10, Hicks12, Kushwaha15}

A situation in which nearly free electrons flow in close proximity to layers of transition-metal oxides in which electron correlations are expected to be strong is highly unusual, and the delafossite metals have already displayed rich magnetotransport and surface state physics, some of which is a direct result of interlayer coupling. \cite{Kim09, Takatsu13, Noh14, Kikugawa16, Moll16,Sunko17,Mazzola17} A further attraction is the simplicity of their bulk Fermi surfaces.  In the Pd- and Pt-based cobaltates and rhodate, a single band crosses the Fermi level, and the strongly two-dimensional electronic structure has facilitated the mapping of the Fermi surface using angle-resolved photoemission spectroscopy (ARPES). \cite{Kushwaha15, Noh14, Noh09, Sobota13, Arnold17} The very long mean free paths have enabled detailed de Haas-van Alphen (dHvA) experiments to complement the ARPES data in PdCoO$_2$, PdRhO$_2$ and PdCrO$_2$ \cite{Hicks12, Arnold17, Ok13, Hicks15} but only preliminary dHvA data have been reported for PtCoO$_2$, for fields close to the crystallographic $c$ axis.\cite{Kushwaha15}  In this article we describe and analyse the results of a comprehensive dHvA rotation study of PtCoO$_2$. Based on these results the weak $c$-axis dispersion is modelled applying similar methods as those previously used for PdCoO$_2$ and PdRhO$_2$. \cite{Hicks12, Arnold17}
Comparing the results to band-structure calculations including a correlation parameter $U$ on the Co site, we obtain a good agreement for strong correlations ($U=6$\,eV), although the $c$-axis warping is slightly overestimated in the calculations.

\begin{figure*}[htb]
	\centering
		\includegraphics[width=1.00\textwidth]{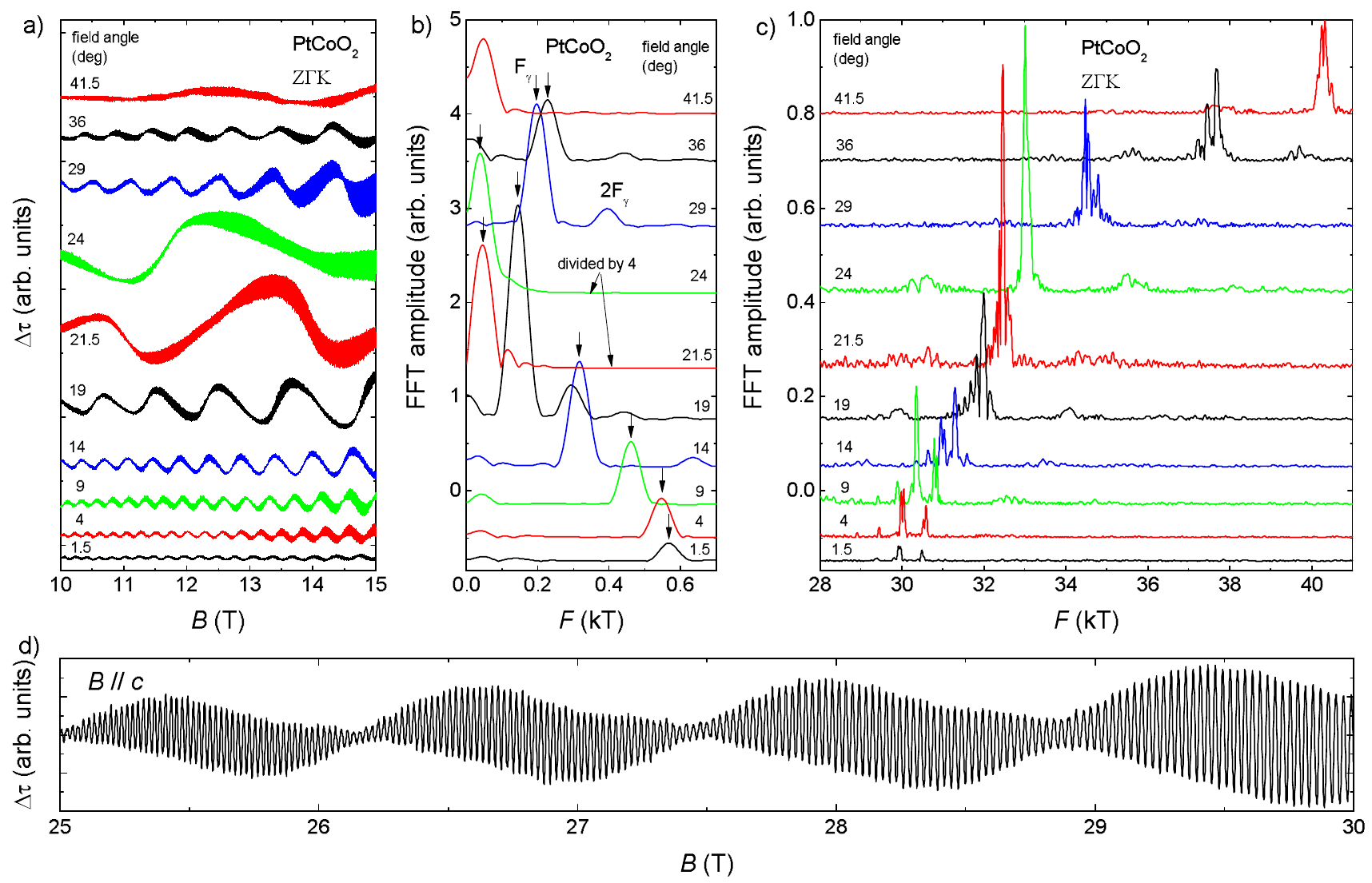}
	\caption[]{a) Exemplary de Haas-van Alphen oscillations in the magnetic torque of the PtCoO$_2$ sample S2 measured at about $700\,\mathrm{mK}$ and in magnetic field applied at indicated angles away from $B||c$ within the $\mathrm{Z}\Gamma\mathrm{K}$ plane. The dHvA signals were obtained by subtracting the paramagnetic background via a second order polynomial from the raw magnetic torque signal. b) and c) Fast Fourier transforms (FFT) were taken between 10 and $15\,\mathrm{T}$ after multiplying the data by a Hanning window. The strong difference frequency $F_\gamma$ is visible both in the oscillatory signal (a) as well as the FFT (b). d) High-field torque oscillations of sample S4 at 370 mK and the magnetic field $B$ around 1 degree from the $c$-axis.}
	\label{fig:DataAndFFT}
\end{figure*}
\section{Experimental Methods}
Crystal growth and characterization of single crystals of PtCoO$_2$ are described in Ref. \onlinecite{Kushwaha17} and references therein. We studied de Haas- van Alphen oscillations of four PtCoO$_2$ crystals, named S1 to S4, by magnetic torque measurements in several laboratories around the globe, with different experimental setups as described below.
Samples S2 and S3 were grown by S. Khim and samples S1 and S4 by H. Takatsu.
  \\Sample S1 was studied at National Institute for Material Science in Japan. The magnetic torque was measured using piezoresistive PRC400 micro-cantilevers in a top-loading type dilution-fridge down to 40 mK in a superconducting magnet and magnetic fields of up to $17.5\,\mathrm{T}$. 
\\Samples S2 and S3 were studied at the Max Planck Institute for Chemical Physics of Solids in Germany. The quantum oscillations of S2 and S3, from the same growth batch, were observed at temperatures between $700\,\mathrm{mK}$ and $4\,\mathrm{K}$ in magnetic fields up to $15\,\mathrm{T}$. The respective sample sizes were approximately $200\times300\times50$\,(\textmu m)$^3$ and $150\times100\times20$\,\,(\textmu m)$^3$.
Experiments on these samples were also performed using 
 piezoresistive PRC400 micro-cantilevers, installed on a MX400 Oxford Instruments dilution refrigerator with a $15/17\,\mathrm{T}$ superconducting magnet and $270^o$ Swedish rotator with an angular accuracy of $\Delta\theta=\pm0.2^\circ$. The magnetometer utilizes a two-stage dc-SQUID (superconducting quantum-interference device) as highly sensitive read-out, offering a torque resolution of $\Delta\tau=2\times10^{-13}\,\mathrm{Nm}$ at lowest temperatures. \cite{Arnold18,Rossel96}
Data were taken at constant temperatures whilst the magnetic field was swept from 15 to $7.5\,\mathrm{T}$ at a rate of $30\,\mathrm{mT/min}$. 
Sample S2 was measured in two orientations so that the field rotated in the crystallographic $\mathrm{Z}\Gamma \mathrm{K}$ and $\mathrm{Z}\Gamma \mathrm{L}$ planes.
\\Sample S4 was studied at the National High Magnetic Field Laboratory in USA. Torque magnetometry was measured by using the capacitive method with a 0.025 mm thick CuBe lever in a $^3$He cryostat and a resistive magnet in fields up to 35 T and a top-loading type dilution fridge in a superconducting magnet in fields up to 16 T.
\\The magnetic field direction in all experiments is given with respect to the crystalline $c$-axis.

Relativistic density functional (DFT) electronic structure calculations were performed using the full-potential FPLO code \cite{Koepernik99, Opahle99} version fplo18.00-52. For the exchange-correlation potential, within the general gradient approximation (GGA), the parametrization of Perdew-Burke-Ernzerhof \cite{Perdew96} was chosen. The spin-orbit coupling (SOC) was treated non-perturbatively solving the four component Kohn-Sham-Dirac equation.\cite{Eschrig} To obtain precise band structure information, the calculations were carried out on a well-converged mesh of 125.000 $k$-points (50x50x50 mesh, 11076 points in the irreducible wedge of the Brillouin zone). The Fermi surface has been calculated on a self-adjusting $k$-mesh to ensure maximum accuracy.
The Coulomb repulsion in the Co-3$d$ shell was simulated in a mean-field way applying the GGA+$U$ approximation in the atomic-limit-flavor. For all calculations, the experimental crystal structure \cite{Prewitt71} was used.


\section{Results and Discussion}

\subsection{Quantum oscillations in PtCoO$_2$}

Typical quantum oscillations in the magnetic torque at different angles for samples S2 and S4 are shown in Fig. \ref{fig:DataAndFFT}. Data from all high-quality samples reported in this paper are mutually consistent. The oscillations consist mainly of three oscillation frequencies. Two close-lying high frequencies, $F_\alpha$ and $F_\beta$ lead to fast oscillations of the signal with varying amplitude due to beating. As expected, the beating frequency is half the difference of the two high frequencies $F_\alpha$ and $F_\beta$. Additionally, we also see strong quantum oscillations with a "slow" frequency $F_\gamma$. As shown in Fig. \ref{fig:AngularDependence} $F_\gamma = F_\beta - F_\alpha$ at all angles. This situation is reminiscent of, but even more pronounced than in, PdCoO$_2$. There, the slow oscillations have been ascribed to magnetic interaction rather than torque interaction, as they also appear in the resistance.\cite{Hicks12, Shoenberg} The same holds for PtCoO$_2$.\cite{Kikugawa16} However, PdRhO$_2$ with a very similar Fermi surface does not show the slow oscillations.\cite{Arnold17} At some angles, additional variations of the oscillation amplitude - slower than the dominant beating caused by $F_\alpha$ and $F_\beta$ - hint to smaller splittings of the main frequencies, also observed in the Fast Fourier transforms (FFT). For example, the oscillations of the torque data at $4\,\deg$ in Fig. \ref{fig:DataAndFFT}a) show an amplitude minimum at around 12 T and concurrently two double peaks appear in the FFT. The reason for this is not understood, but nevertheless, two main peaks could always be identified. Overall, the fast quantum oscillation frequencies follow a $1/\cos{\theta}$ angular dependence 
consistent with a cylindrical quasi-two-dimensional Fermi surface as also observed in PdCoO$_2$ and PdRhO$_2$.\cite{Hicks12,Arnold17}

\subsection{Properties for $B \parallel c$}

For $B \parallel c$ the main frequencies are $F_\alpha=29.92\,\mathrm{kT}$ and $F_\beta=30.49\,\mathrm{kT}$ with a difference frequency of $F_\gamma = 0.57\,\mathrm{kT}$.

\begin{figure}[tb]
	\centering
		\includegraphics[width=0.95\columnwidth]{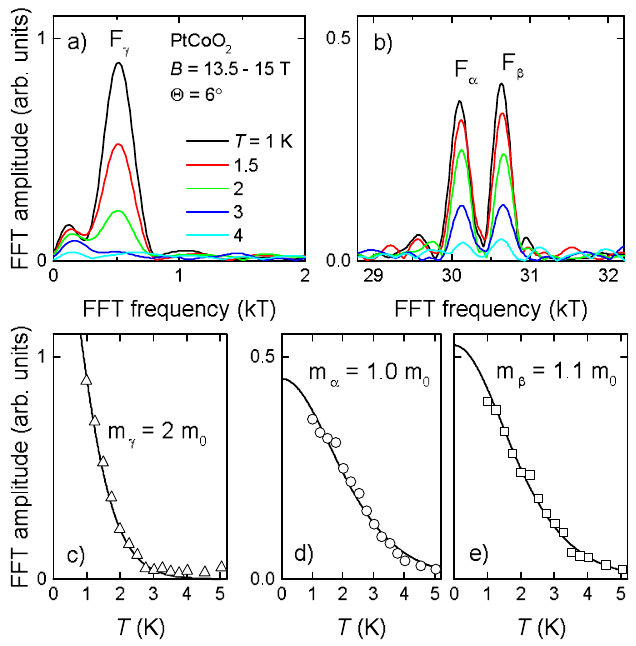}
	\caption[]{a) and b) FFT amplitude for fields in the range $13.5\,\mathrm{T}$ to 15$\,\mathrm{T}$ at a magnetic field angle of $6^\circ$ off the $c$-axis in the $\mathrm{Z}\Gamma\mathrm{L}$ plane and temperatures as indicated.  c) - e) Temperature dependence of the oscillation peak amplitudes (markers). Lifshitz-Kosevich fits (lines) give the effective masses as indicated within an uncertainty of 10\%, where m$_0$ is the free electron mass.
	}
	\label{fig:EffectiveMasses-Hanning}
\end{figure}

The measured mean quantum-oscillation frequency of $\overline{F}_0=30.20 \pm 0.01\,\mathrm{kT}$ results in a Fermi-surface cross section of $S_\mathrm{ext}=2.883\,$\AA$^{-2}$. The cross section multiplied by the height of the Brillouin zone gives the Fermi-surface volume, which is, according to Luttinger's theorem, directly related to the carrier density. Given the lattice constants of $a=2.82\,$\AA~and $c=17.808\,$\AA ,\cite{Kushwaha15} the experimental Fermi-surface cross section is only 0.6 \% larger than the cross section from a half-filled metal and hence corresponds to a Luttinger count of 1.006. A half-filled metal is expected when each Pt atom contributes one electron to the conduction band. Using a slightly larger lattice parameter from Ref. \onlinecite{Prewitt71} of $a=2.83\,$\AA~(the value used for the DFT calculations below), the Fermi-surface cross section is 1.4 \% larger than the half-filled value. We note, however, that these lattice parameters are based on room-temperature x-ray diffraction, and that thermal contraction likely means that our true Luttinger count is close to 1.

By measuring the temperature dependence of the quantum-oscillation amplitude of sample S2, we extract the cyclotron masses for the three dominant frequencies close to $B\parallel c$, as shown in Fig. \ref{fig:EffectiveMasses-Hanning}. Fitting the data by the Lifshitz-Kosevich temperature reduction term yields the cyclotron masses $m_\alpha = (1.0 \pm 0.1)\,m_\mathrm{0}$, $m_\beta = (1.1 \pm 0.1)\,m_\mathrm{0}$ and $m_\gamma = (2 \pm 0.2)\,m_\mathrm{0}$ where $m_\mathrm{0}$ is the free electron mass. Within error bars, this result is independent of field-window size and reasonable variations of the temperature window of the fit. These values are slightly lower than effective masses from ARPES,\cite{Kushwaha15} where a mean effective mass of $m = 1.18\,m_0$ was found. ARPES results show that the effective mass is $k$-dependent within the plane, dominated by the variation of $k_F$ in the hexagonally shaped Fermi-surface cross section. A possible $k$ dependence also along the $c$ direction might explain the difference to the results from quantum oscillations which give mean effective masses around orbits at different heights in the Brillouin zone. 

Combining the measured quantum-oscillation frequencies with the cyclotron masses, we obtain the averaged Fermi velocities of $v_{\mathrm{F}} = \hbar k_\mathrm{F}/m^\star$ $v_{\mathrm{F}\alpha}=(1.1 \pm 0.1)\times10^6\,\mathrm{m/s}$ and $v_{\mathrm{F}\beta}=(1.0 \pm 0.1)\times10^6\,\mathrm{m/s}$, which are as high as in simple metals \cite{AshcroftMermin} and are in reasonable agreement with the direction-independent Fermi velocity from ARPES of $(0.89 \pm 0.09)\times10^6\,\mathrm{m/s}$.

At higher fields (25 to 30 T), data on sample S4 give effective masses $m_\alpha = (1.15 \pm 0.1)\,m_\mathrm{0}$, $m_\beta = (1.23 \pm 0.1)\,m_\mathrm{0}$ and $m_\gamma = (2 \pm 0.2)\,m_\mathrm{0}$ that are slightly higher than those determined for sample S2 but in agreement with the low-field values within the respective uncertainties.

A very rough estimate of the Dingle temperature was obtained for data of sample S2 near $B\parallel c$ by cutting the total inverse field range in small equal windows (still ranging over several beating periods), taking the FFT of the small windows and extracting thereby the field dependence of the amplitude (height) of the FFT peaks. We find $T_\mathrm{D} \approx 3.5 \pm 1.5\,\mathrm{K}$ for both main frequencies. The careful error bar here includes the uncertainty of the effective mass as well as variations of $T_\mathrm{D}$ when the field-window size is changed. Using the relations $T_\mathrm{D}=\frac{\hbar}{2\pi k_\mathrm{B}\tau}$ and $v_\mathrm{F}=\hbar k_\mathrm{F} = l_0/\tau$ we obtain a mean free path $l_0$ of roughly $350 \pm 150\,\mathrm{nm}$. 
It is worth noting that the mean free path obtained here is at least an order of magnitude smaller than the one extracted from residual resistivity values on samples of the same batch \cite{Kushwaha15} of around 5\,\textmu m. The large mean free path in resistivity might be related with the hexagonal shape of the Fermi surface with large flat portions. There, small-angle scattering does not change the Fermi velocity and hence is completely ineffective for resistivity. However, all scattering events lead to Landau level broadening and the subsequent suppression of the quantum oscillation amplitude.

\subsection{Angular dependence of quantum oscillations}
\begin{figure}[htb]
	\centering
		\includegraphics[width=0.9\columnwidth]{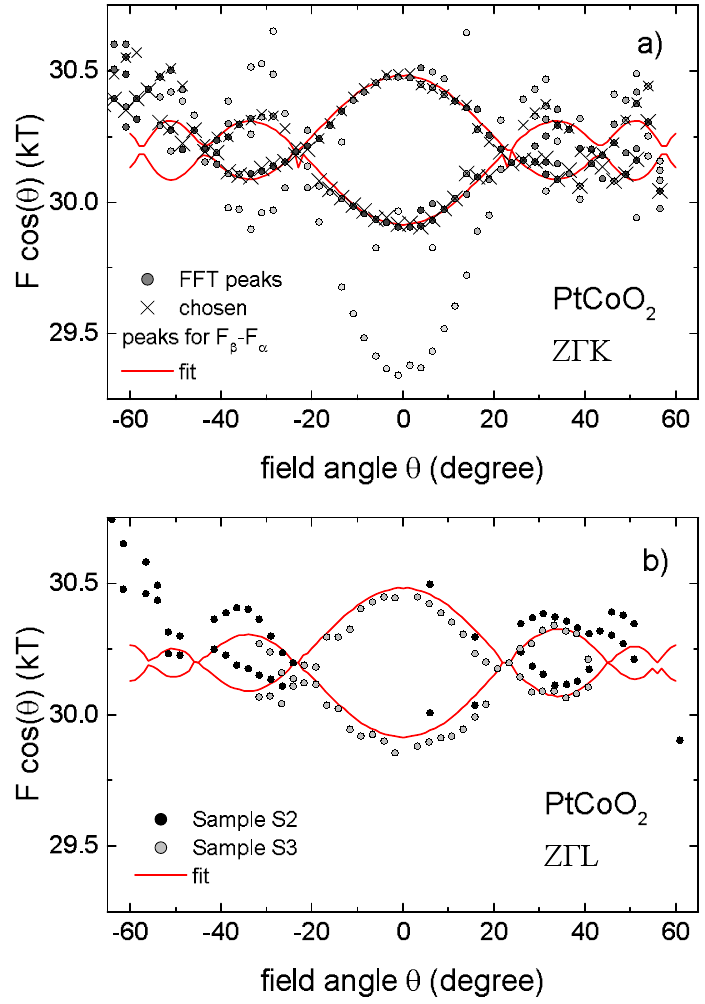}
	\caption[]{
	Angular dependence of the main quantum oscillation frequencies multiplied by $\cos{\theta}$ for the two tilting planes $\mathrm{Z}\Gamma \mathrm{K}$ (a) and $\mathrm{Z}\Gamma \mathrm{L}$ (b). The experimental values are to be compared with the angular dependence of a Fermi surface with the warping parameters given in Fig. \ref{fig:TopographyFigure}. In a) the peaks in the FFT spectrum taken on data between 10 and 15 T of sample S2 (dots) were found numerically with angle-dependent conditions on peak prominence and relative height to the maximum peak height of $\alpha$ or $\beta$ peaks. The relative peak height is also given by a grey shade of the dots with black being the maximum height at that angle. Peak splittings and side peaks with $F = F_\alpha - F_\gamma$ or $F = F_\beta + F_\gamma$ also appear as visible in Fig. \ref{fig:DataAndFFT}c. Crosses indicate the chosen peaks for $\alpha$ and $\beta$ when the peak selection was done by hand. These frequency values were used for the calculation of the difference frequency $F_\beta - F_\alpha$ in Fig. \ref{fig:AngularDependence}. Some of these peaks slipped through the numerical selection when the peak prominence was too small. 
	\\ In b) only the peaks selected by hand are shown without amplitude shading. The black (grey) dots give the angular dependence of sample S2 (S3), respectively.
	}
	\label{fig:AngularDependenceRaw}
\end{figure}


\begin{figure}[tb]
	\centering
		\includegraphics[width=0.9\columnwidth]{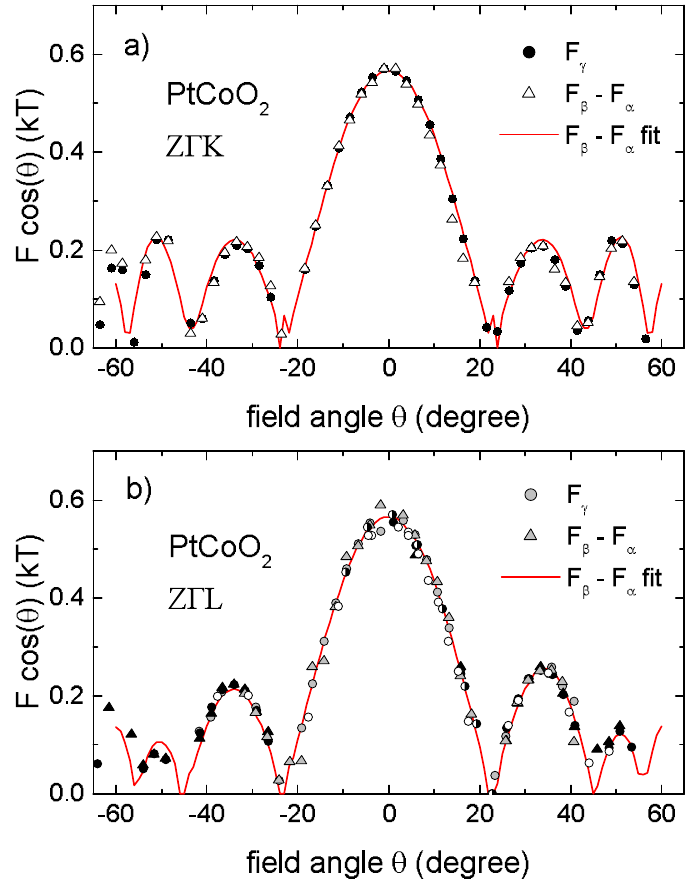}
	\caption[]{
	Angular dependence of the quantum oscillation frequency $F_\gamma$ (dots) and the difference $F_\beta - F_\alpha$ (triangles) in PtCoO$_2$ for magnetic fields within the $\mathrm{Z}\Gamma \mathrm{K}$ (a) and $\mathrm{Z}\Gamma\mathrm{L}$ plane (b) compared with the values from the approximated Fermi surface of Fig. \ref{fig:TopographyFigure}. In a) experimental data are from sample S2, in b) black symbols are from sample S2, grey symbols from sample S3, white symbols from sample S4 and half black symbols from sample S1.
	}
	\label{fig:AngularDependence}
\end{figure}

The main quantum oscillation frequencies vary with angle around their mean value by $\overline{F}_0/\cos{\theta}$, i.e. a constant in Fig. \ref{fig:AngularDependenceRaw} expected for a perfectly cylindrical Fermi surface. Note that the jumps/scatter in the angular dependence are due to small uncertainties in angle resulting in rather large variations of the factor $1/\cos{\theta}$.\cite{Arnold17}  In addition, a slight lag of our rotator is evidenced by a departure from the $1/\cos{\theta}$ behavior at large angles. In Fig. \ref{fig:AngularDependenceRaw}b) the mean value $\overline{F}_0 = 1/2 (F_\alpha + F_\beta)$ is slightly lower in sample S3 than in sample S2, perhaps indicating tiny variations of less than 0.1\% in electron count between samples.
Deviations of the quantum-oscillation frequency from the mean frequency (imaginary horizontal line in Fig. \ref{fig:AngularDependenceRaw}) indicate a departure of the Fermi surface from an ideal cylindrical shape -- a warping of the Fermi surface -- leading to a splitting of the extremal cyclotron orbits into a neck and a belly orbit. At certain angles of the magnetic field, so-called Yamaji angles, the splitting reduces to zero when all cyclotron orbits inclose the same cross section. In principle, some warping parameters (such as a double $c$-axis warping -- $k_{0,2}$ below and in Fig. \ref{fig:TopographyFigure} -- or strong \grqq{bumps}\grqq\,\,as in Fig. \ref{fig:HR2} and \ref{fig:HR3} for small $U$) can cause a departure of the mean frequency from the $1/\cos{\theta}$ dependence. However, because of the scatter in the raw data, we cannot confirm the presence of small contributions of this type.

The angular dependence of $F_\gamma$ is given in Fig. \ref{fig:AngularDependence}. It displays exactly the same behavior as the difference of the two main frequencies $F_\beta - F_\alpha$, supporting that  $F_\gamma$ stems from magnetic interaction of $F_\beta$ and $F_\alpha$. Compared to the raw data, the problems of angle uncertainty cancel out here and a smooth curve is recovered (triangular symbols in Fig. \ref{fig:AngularDependence}). In this presentation, information on warping causing a departure of the mean frequency from the $1/\cos{\theta}$ dependence mentioned above is lost.

The difference frequency $F_\gamma$ is reminiscent of the "slow" oscillations observed in quasi two-dimensional organic conductors \cite{Kartsovnik02} although there are some differences. In the organic conductors, the slow oscillations appear only in the interlayer conductivity and are absent in the magnetic torque. They are in good agreement with a theory where fast oscillations in the density of states are mixing with fast oscillations of the chemical potential or Dingle temperature. This leads to interference effects and hence combination frequencies \cite{Grigoriev03} in the conductivity. A similar mixing can also lead to slow oscillations in the magnetization as a form of magnetic interaction.\cite{Shoenberg, Grigoriev01}
The reason why slow oscillations in the magnetization appear only in the delafossites might be the very high purity of these systems but we cannot rule out other origins for the observations. The absence of the slow frequency in PdRhO$_2$ remains a mystery, and presents a challenge to interpretations.

The degree of deviation from two-dimensionality can be estimated via the $c$-axis hopping parameter $t_\perp$.
It is given by $2t_\perp=E_\mathrm{F_\beta}-E_\mathrm{F_\alpha}$ using $E_\mathrm{F} = \hbar^2 k_\mathrm{F}^2/2\overline{m}^\star$ with Planck's constant $\hbar$ and the mean effective mass $\overline{m}^\star = 1.05\, m_0$ in units of the free electron mass $m_0$. The Fermi energies $E_\mathrm{F}$ are then directly related to the quantum oscillation frequencies by the Onsager relation $F=\frac{\hbar}{2\pi e} A_\mathrm{ext}$ with the extremal Fermi-surface area in reciprocal space $A_\mathrm{ext}=\pi k_\mathrm{F}^2$ and the electron charge $e$. Hence we obtain $t_\perp= \frac{\hbar e}{\overline{m}^\star}\Delta F = 31$\,meV. We can expect a two-dimensional behavior when the Landau-level splitting $\hbar \omega_c$ is larger than the dispersion along $c$. We calculate $\hbar \omega_c = \hbar eB/m^{\star}$ at 15 T with the effective masses of $m^\star \approx m_\mathrm{e}$ and obtain $\hbar \omega_c = 1.8$\,meV, a factor of 17 smaller than $t_\perp$. This implies that a field of 290 T would be needed to reach perfect two-dimensionality in this compound.

\subsection{Fermi surface warping parameters}

\begin{figure}[tb]
	\centering
		\includegraphics[width=0.8\columnwidth]{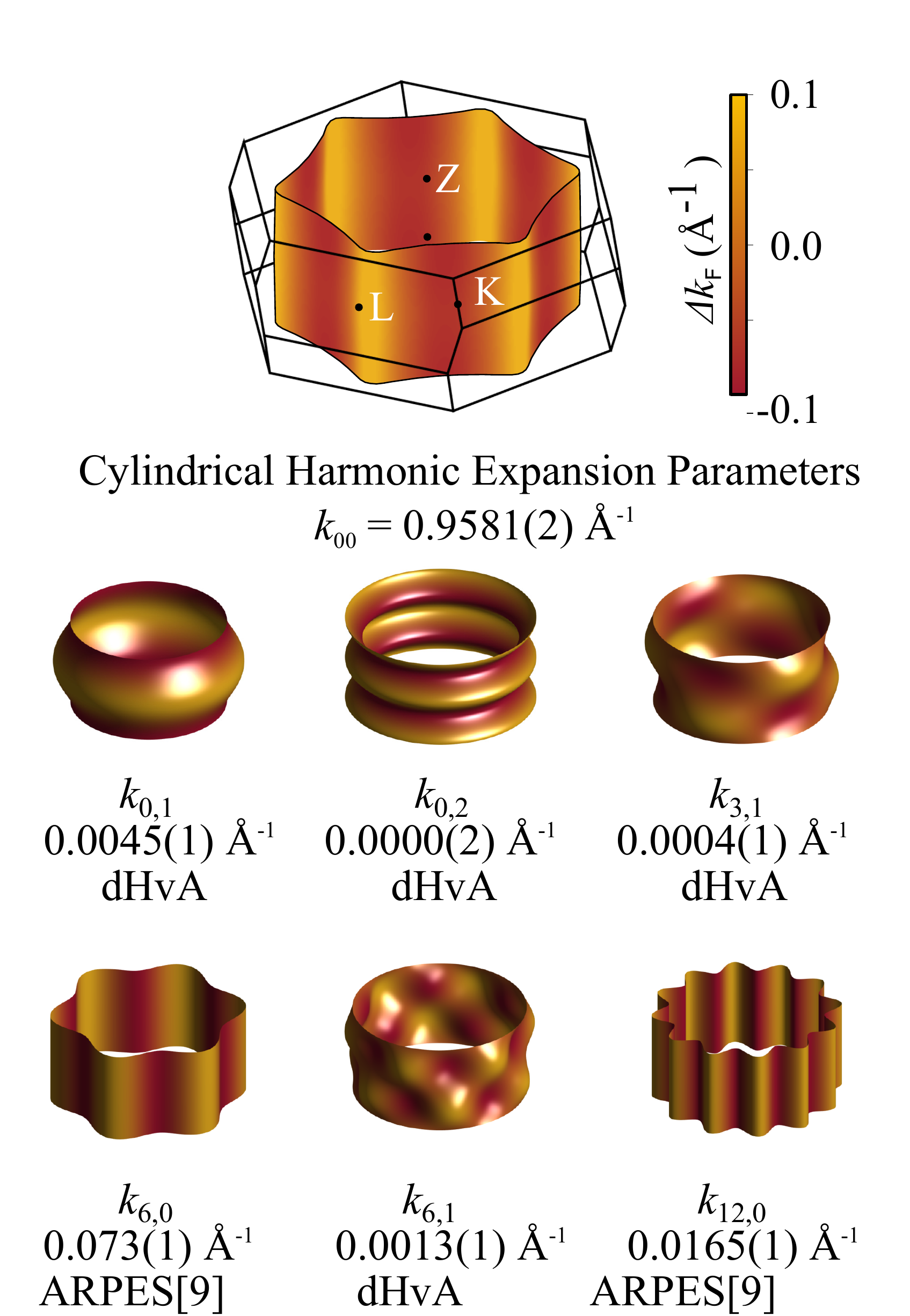}
	\caption[]{Reconstructed Fermi surface topography and cylindrical harmonic expansion parameters of PtCoO$_2$. }
	\label{fig:TopographyFigure}
\end{figure}

We extract the Fermi-surface warping parameters from the deviation of the two high frequencies from the $1/\cos{\theta}$ angular dependence. To model the experimental angular dependence a cylindrical harmonic expansion is used \cite{Arnold17, Hicks12, Bergemann00, Bergemann03}
\begin{eqnarray}
k_\mathrm{F}=\sum_{\mu,\nu\geq0}{k_{\mu,\nu}\cos(\nu\kappa)\cos(\mu\phi)},
\label{eqn:warping}
\end{eqnarray}
where $\kappa=c^*k_z$ is the reduced $z$-coordinate and $\phi$ the azimuthal angle.\cite{Arnold17} Note that the interlayer spacing $c^*$ is a third of the $c$-axis lattice constant. The hexagonal lattice symmetry and $R\overline{3}m(D^5_{3d})$ space group limit the allowed $k_{\mu,\nu}$ to $(\mu,\nu) \in \{(0,0);(0,1);(0,2);(0,3);(3,1);(6,0);(12,0)\}$ and higher order terms. The  $k_{\mu\nu}$  given in Fig. \ref{fig:TopographyFigure} can reproduce the experimental angular dependence in both field-tilting planes (red lines in Fig. \ref{fig:AngularDependenceRaw} and \ref{fig:AngularDependence}). The in-plane parameters $k_{6,0}$ and $k_{12,0}$ were obtained from the Fermi surface shape determined by ARPES.\cite{Kushwaha15}

Let us compare the Fermi surface parameters of PtCoO$_2$ with PdCoO$_2$.\cite{Hicks12} Both have very similar lattice constants (the in-plane lattice constant $a$ is 2.82\,\AA~in PtCoO$_2$ compared to 2.83\,\AA~in PdCoO$_2$ and the inter-plane lattice constant $c$ is 0.4\% larger in PtCoO$_2$ with 17.808\,\AA~compared to 17.73\,\AA.
On the one hand, PtCoO$_2$ has a much smaller splitting of the main frequency, leading to a $k_{0,1}$ smaller by a factor of 2.4 than in PdCoO$_2$. This points to little direct overlap of the Pt $5d$ orbitals along the $c$-direction, the value being similar to PdRhO$_2$.\cite{Arnold17} On the other hand, $k_{3,1}$ is of the same order in both compounds (and an order of magnitude smaller than in PdRhO$_2$). In PdCoO$_2$, DFT calculations without correlations on the Co site showed a too high hybridisation between the conduction electrons and the Co electrons and consequently a high $k_{3,1}$ (similar to the \grqq bumps\grqq\,\,in Fig. \ref{fig:HR2} for $U<4$\,eV). Introducing a correlation parameter $U$ on the Co site of around 3 eV reduced $k_{3,1}$ to a value similar to the experimental one. 

\subsection{Band-structure calculations in PtCoO$_2$}

\begin{figure}[tb]
	\centering
		\includegraphics[width=0.8\columnwidth]{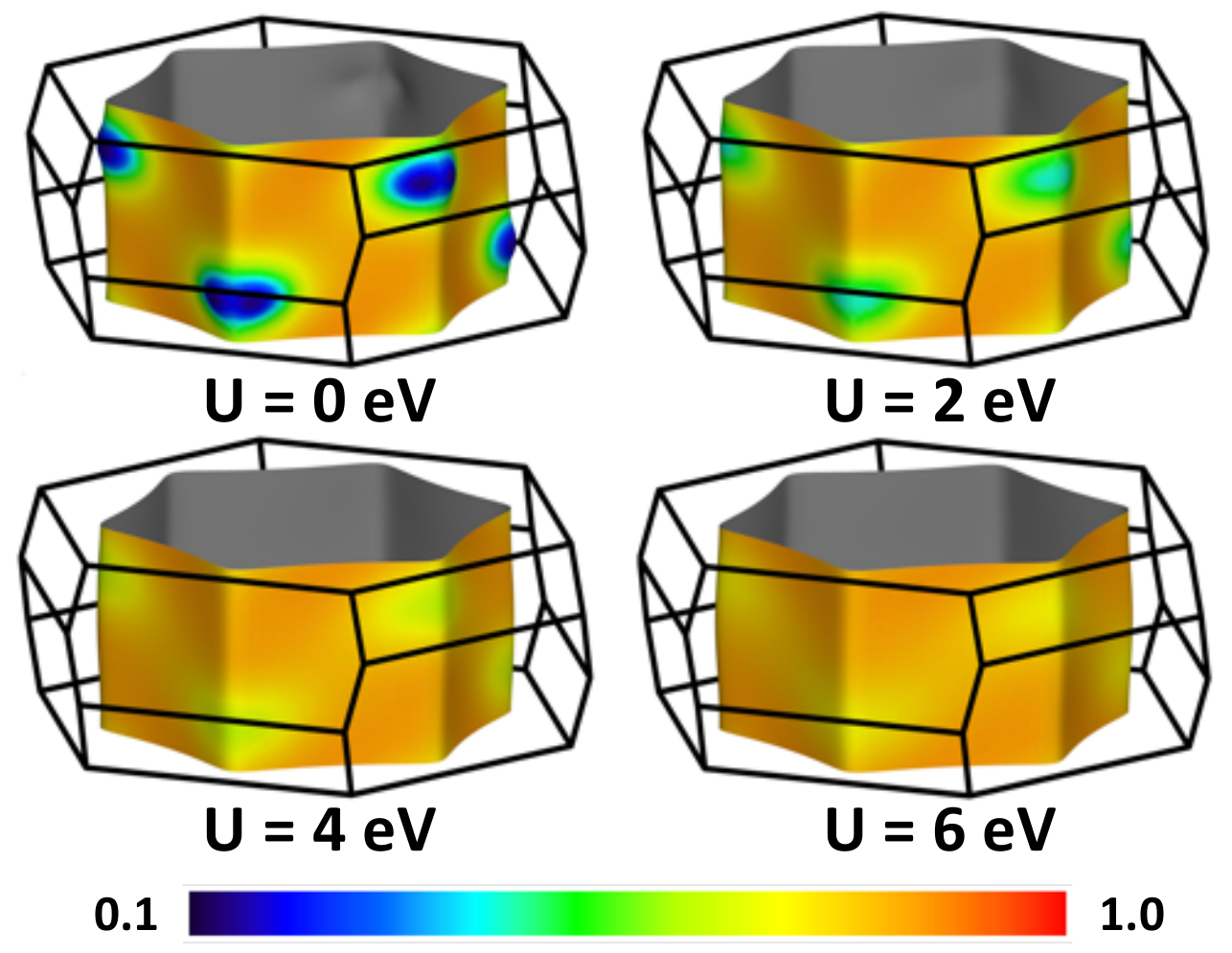}
	\caption[]{Impact of the Coulomb parameter $U$, used to describe the strong correlations in the Co-$3d$ orbitals, on the DFT-calculated  Fermi surface. The related Fermi velocities are shown by colour coding (in $10^{-6}$\,m/s). Without $U$, the Fermi surface exhibits \grqq{bumps}\grqq\,\,incompatible with the experimental observations. With increasing $U$, the \grqq{bumps}\grqq\,\,disappear together with the related strong inhomogeneity of the Fermi velocities.}
	\label{fig:HR2}
\end{figure}

\begin{figure}[tb]
	\centering
		\includegraphics[width=1\columnwidth]{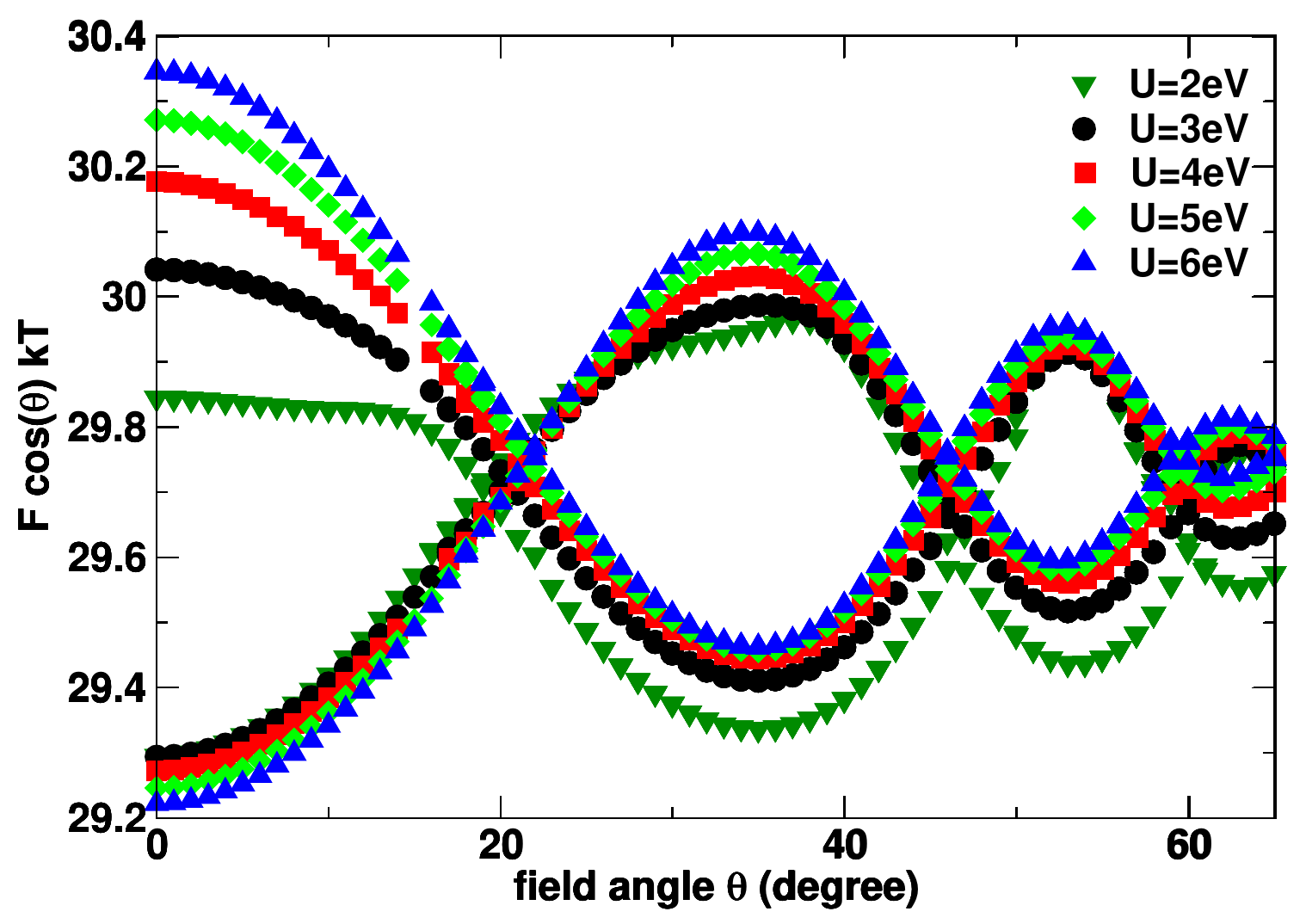}
	\caption[]{Angular dependence of the DFT-calculated Fermi surface cross sections, multiplied by $\cos{(\theta)}$ for the Z$\Gamma$L tilting plane. The frequencies depend significantly on the Coulomb parameter $U$ that is used to describe the strong correlations in the Co-$3d$ orbitals. }
	\label{fig:HR1}
\end{figure}

\begin{figure}[tb]
	\centering
		\includegraphics[width=.8\columnwidth]{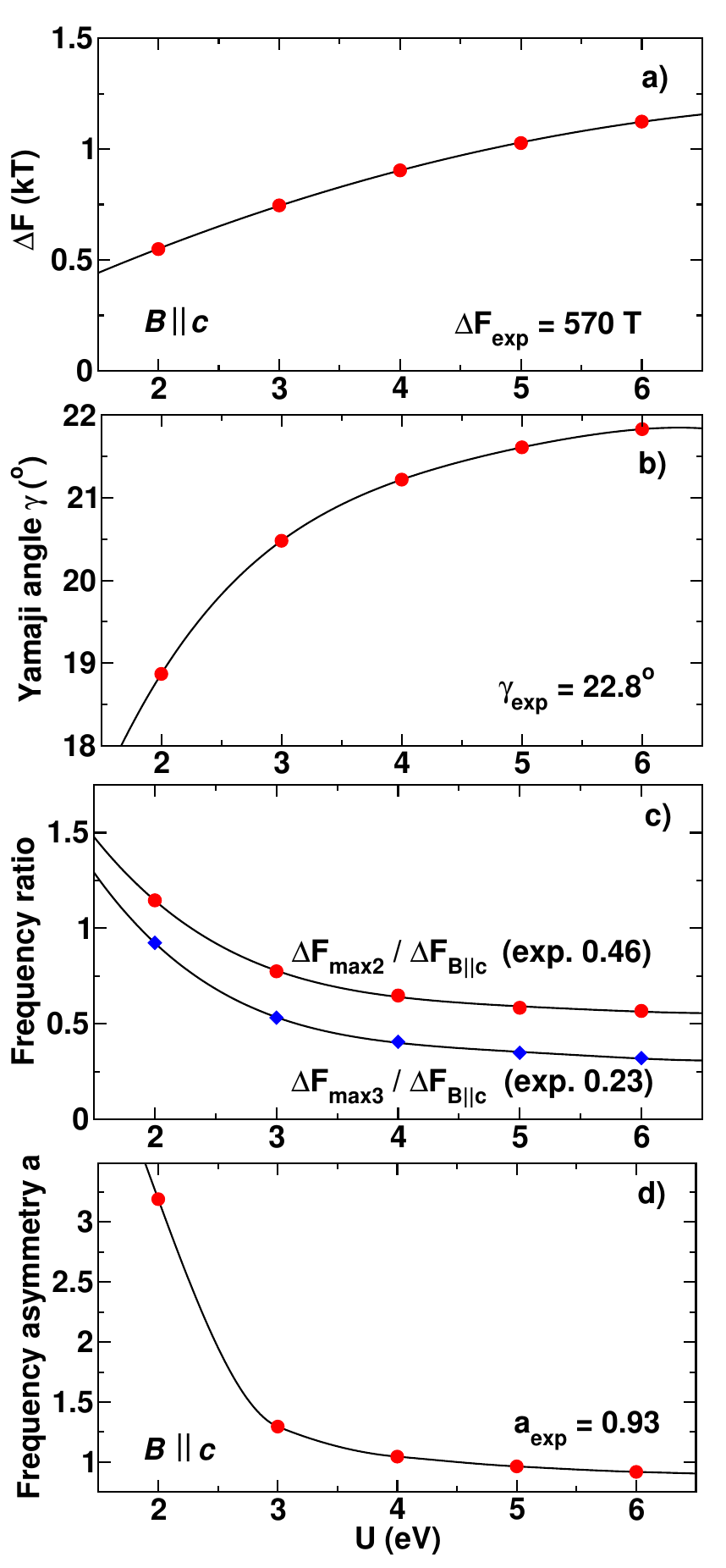}
	\caption[]{Dependence of the calculated Fermi surface properties on the Coulomb parameter $U$. Lines are guides to the eye. The respective experimental results are provided in the individual panels for comparison. Panel a) shows the frequency split for the magnetic field $B \parallel c$. The Yamaji angle is shown in b). Panel c) provides the ratios for the 2nd and 3rd maximum frequency split with respect to the split for $B \parallel c$ from panel a). Note that the 2nd maximum split $\Delta F_\mathrm{max2}$ is at around 35\,$\deg$ and the 3rd maximum split $\Delta F_\mathrm{max3}$ at around 53\,$\deg$ in the Z$\Gamma$L tilting plane, compare Fig.\,\ref{fig:HR1}. Panel d) provides the asymmetry $a$ of the latter frequency split with respect to the frequency at the respective Yamaji angle. The definition of $a$ is given in the main text.}
	\label{fig:HR3}
\end{figure}

Density functional calculations, applying GGA for the exchange correlation potential and including spin orbit coupling describe the Fermi surface (FS) of PtCoO$_2$ rather well.\cite{Kushwaha17} The result is a single band crossing the Fermi energy which is predominantly derived from Pt 5$d$ states. This band shows a weak dispersion along the $z$ direction, resulting in a Fermi surface shaped like a hexagonal cylinder (see Fig.\,\ref{fig:HR2}). However, this band shows also a sizeable admixture from Co 3$d$ states, leading to \grqq bumps\grqq\,\,at the edges of the hexagonal cylinder-shaped Fermi surface (see Fig.\,\ref{fig:HR2}, upper left panel). These \grqq bumps\grqq\,\,in the calculations are characterised by their low Fermi velocities due to the small band width of the related Co states. In an ARPES experiment, these \grqq bumps\grqq\,\,were not observed,\cite{Kushwaha17} leading to the conclusion that they are artefacts and a result of the underestimation of the strong correlations of the Co 3$d$ states in the insulating CoO$_2$ layer by the DFT calculations. Applying DFT\,+\,$U$ as a mean field description for the strong Coulomb repulsion in these orbitals, the experimental shape of the Fermi surface could be reproduced within the resolution of the ARPES experiment for a moderate value of $U = 4$\,eV.

In contrast to the ARPES experiment, which provides a two-dimensional projection of the Fermi surface, angular dependent dHvA measurements yield a highly precise scan of the three-dimensional Fermi surface. This enables a more detailed study of the role of correlations in the CoO$_2$ layer and their influence on the Pt-derived Fermi surface. Figure\,\ref{fig:HR1} shows the calculated angular dependent dHvA frequencies as a function of the value of $U$ for the Co 3$d$ states. Overall, the curves reproduce the experimental data very well (see Fig.\,\ref{fig:AngularDependenceRaw}).  At first glance, they are rather similar in absolute frequency values, showing deviations from each other of the order of 1\%, only.  This is expected since the averaged frequency is essentially fixed by the choice of lattice parameters due to the half-filled nature of the hexagonal cylindrical Fermi surface. However, the plot shows a significant dependence of the frequency splits and the Yamaji angle on the $U$ parameter, which will be investigated now.

Comparing the shape of the calculated angular dependencies with the experiment (see Fig.\,\ref{fig:AngularDependenceRaw}) the agreement for small $U$ values (less than 3\,eV) is rather poor. Especially, the flattening of the upper frequency at small angles and near the second maximum at around $35\,\deg$ for $U=2$\,eV is not reflected in the experiment. This is in line with the still significant Co $3d$ contribution at the Fermi surface (see Fig.\,\ref{fig:HR2}, upper right panel with $U=2$\,eV). The details of the Fermi surface warping depend on the choice of $U$ as presented in Fig.\,\ref{fig:HR3}\,\,showing (a) the frequency split for $B \parallel c$, (b) the Yamaji angle (the angle where the frequency split becomes zero), (c) the ratio of the split maxima at different angles and (d) the split asymmetry $a$ for $B \parallel c$ ($a = (F_Y - F_\alpha) / (F_\beta - F_Y)$ where $F_\alpha$ and $F_\beta$ are the lower and the upper frequency for $B \parallel c$, respectively, and $F_Y$ is the frequency at the Yamaji angle) together with the respective experimental values. Essentially, all plots show a saturation for $U > 6$\,eV. The latter three yield good agreement with the experiment for the larger $U$ values. In contrast, the frequency split for $B \parallel c$ is overestimated by a factor of about 2 for $U=6$\,eV. This overestimate can be directly linked to the dispersion due to the Pt interlayer coupling $t_\perp$. From our calculation, directly estimating the dispersion of the half-filled band at the Fermi level, we extract $t_\perp = 55$\,meV (for $U=6$\,eV) compared to $t_\perp = 31$\,meV from the experiment. The origin of this DFT-overestimate is unclear, but similar effects have been observed for many other low-dimensional compounds for the couplings in the \grqq weak\grqq\,\,directions.\cite{HR97,DK08}

As result of our DFT calculations, we find a good agreement of the calculated and the experimental Fermi surface when we include strong correlations for the Co 3$d$ states ($U=6$\,eV). The assignment of a larger $U$ value compared to the ARPES description with $U=4$\,eV \cite{Kushwaha17} results from a greater sensitivity of the dHvA data to the choice of $U$ and points to a strongly correlated regime in the CoO$_2$ layers of PtCoO$_2$.

Independent of detail, our analysis demonstrates that strong correlations in the transition-metal layers play an important role in determining the properties not just when the transition-metal ion is magnetic \cite{Ok13, Hicks15, Sunko18} but also when it is non-magnetic like Co in PtCoO$_2$.

\section{Summary and Conclusion}
This quantum-oscillation study combined with band-structure calculations gives detailed information on the Fermi surface of PtCoO$_2$. Like the sister compounds PdCoO$_2$ and PdRhO$_2$, it is a half-filled metal with a single quasi-two-dimensional Fermi surface with hexagonal cross-section, in the case of PtCoO$_2$ formed by Pt states. The effective masses are near the free electron value. By following the angle dependence of the two main quantum-oscillation frequencies, the warping parameters of the Fermi surface along the out-of-plane direction were determined. The warping along the $c$-axis is small, but its details enable an investigation of correlation effects in the CoO$_2$ layers on the Fermi surface by comparing the experimental result with state-of-the art electronic structure calculations. Many of the specific warping details such as the Yamaji angle and the ratio between first, second and third frequency-split maximum with increasing angle, can be reproduced quantitatively by the calculations, when a strong Coulomb repulsion of $U=6$\,eV is included in the CoO$_2$ layers. The high precision and accuracy of dHvA data, combined with the effects of interlayer coupling, thus allow this study of the Fermi surface to yield information on the physics of strong correlations in the insulating CoO$_2$ spacer layers.

\section{Acknowledgments}
The authors would like to acknowledge the financial support from the Max-Planck Society. 
EH and MN acknowledge support from DFG through the project TRR80: From electronic correlations to functionality.
We are grateful to late Professor James S. Brooks (NHMFL, Florida State University) for his continuous support and encouragement during this work. 
This work is supported by JSPS KAKENHI (No. 18K04715). 
A portion of this work was performed at the National High Magnetic Field Laboratory, which is supported by National Science Foundation Cooperative Agreement No. DMR-1157490 and the State of Florida.

\bibliographystyle{unsrtnat}

\end{document}